\documentclass[11pt]{article}
\usepackage[english]{babel}
\usepackage{graphicx}
\usepackage{amsmath}
\usepackage{amssymb}
\usepackage{amsxtra}
\usepackage{float}

\title{Anihelium flux from antimatter globular cluster}
\author{M.Yu. Khlopov$^{1,2,3}$, A.O. Kirichenko$^{2}$, A.G. Mayorov$^{2}$\\
$^{1}$ Institute of Physics, Southern Federal University\\ Stachki 194 Rostov on Don 344090, Russia\\
$^{2}$ National Research Nuclear University MEPhI \\(Moscow Engineering Physics Institute),\\ 115409 Moscow, Russia\\e-mail aokirichenko@yandex.ru\\
$^{3}$  Université de Paris, CNRS, Astroparticule et Cosmologie,\\ F-75013 Paris, France}

\begin{document}
\maketitle

\begin{abstract}

 Macroscopic cosmic antimatter objects are predicted in baryon asymmetrical Universe in the models of strongly nonhomogeneous baryosynthesis. We test the hypothesis of the existence of an old globular cluster of anti-stars in the galactic halo by evaluating the flux of helium anti-nuclei in galactic cosmic rays. Due to the symmetry of matter and antimatter we assume that the antimatter cluster evolves in a similar way as a matter cluster.
The energy density of antiparticles in galactic cosmic rays from antimatter globular cluster is estimated.
 We propose a method for the propagation of a flux of antinuclei in a galactic magnetic field from the globular cluster of antistars in the Galaxy.

\end{abstract}



\section{Introduction}\label{s:intro}
At the end of the 1920s, Paul Dirac predicted the existence of antiparticles — that is new particles, which are opposite in sign of electric, baryonic, lepton  and other charges of already known particles \cite{link1}.
Antimatter is detected in cosmic rays. According to the modern concepts it has three possible nature of origin:
\begin{itemize}
    \item Primordial antimatter. It could be created in the early Universe as the reflection of nonhomogeneous baryosynthesis \cite{newBook,DolgovAM}, evolve in antimatter domains and now it can exist in the form of macroscopic antimatter objects like globular clusters of antimatter stars \cite{newBook}.
     \item Secondary antimatter. It is formed as a result of the collision of the nuclear component of cosmic rays with interstellar gas or with a supernova shell \cite{Oliva}.
     \item Antimatter from exotic sources like evaporation of primordial black holes or the decay/annihilation of hypothetical particles of dark matter \cite{newBook}.
\end{itemize}
 According to \cite{link2}, such object can be present in the Galaxy in the form of a globular cluster of antimatter stars. The prediction\cite{link2} assumes similarity in the properties of antimatter and matter globular clusters. Based on this similarity we consider here possibilities to test the hypothesis of antimatter globular cluster in searches for antihelium component of cosmic rays.
Our approach is aimed to specify the predictions of this hypothesis with the account of realistic description of the production and propagation of cosmic antihelium fluxes in the Galaxy. 

\section{Primordial antimatter}
The baryon asymmetry of the Universe is the observed predominance of matter over antimatter in the visible part of the Universe.
Explaining the origin of the baryon asymmetry of the Universe is one of the key problems of modern cosmology and physics of elementary particles.
A. D. Sakharov\cite{link3} and V.A. Kuzmin\cite{link4} formulated the necessary conditions for bariosynthesis in the early Universe:

1. Asymmetry between particles and antiparticles as a violation of charge C- and combined CP-symmetry. 

2. Violation of the law of conservation of baryon charge. 

3. Violation of local thermodynamic equilibrium. 

On the other hand, it was shown in \cite{link5} - \cite{link8} that almost all existing
mechanisms of baryosynthesis allow the existence of domains with an excess of antimatter, if baryosynthesis is strongly nonhomogeneous. The size of domains depends on the details of the considered mechanisms and can be both small and reaching the size of a Metagalaxy.

The macroscopic region of antimatter with an excess of antibaryons at the same temperature and density evolves in the same way as ordinary matter of macroscopic size. Experiments on accelerators synthesizing antimatter show that the properties of particles and antiparticles coincide, except of the small effect of CP-violation \cite{link9} .

An astronomical object smaller than a globular cluster cannot be formed surrounded by matter during cosmological evolution\cite{link10}. With smaller sizes, antimatter would annihilate with matter before the Galaxy formation. The larger size of domains is constrained by the observed fluxes  of gamma radiation.

Globular clusters of antistars could form during the formation of the Galaxy and remain in its halo by now.

Cosmic ray fluxes of antinuclei are the profound signature of antimatter stars and provide the probe of their existence.

\section{Secondary antimatter}  
\begin{figure}[htp]
\centering
\includegraphics[scale=0.7]{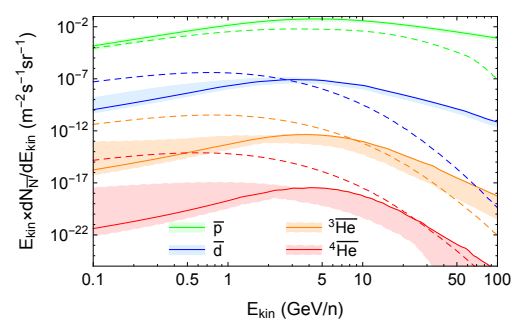}\\
\caption{\small Upper limits for of secondary \cite{link13} antihelium, antiproton, antideuteron together with previous results.}
\label{flux}
\end{figure}

The detected fluxes of cosmic antiparticles are formed as a result of collisions of high-energy nuclear component of cosmic rays with interstellar gas. Study of the processes of antiproton and light antinuclei production at accelerators made it possible to determine the cross section of these processes. The data obtained were used to predict the cross sections for heavier nuclei.
This analysis (figure \ref{flux}) shows that detection of $ {\overline{He}}^3$, $ {\overline{He}}^4$  at level of sensitivity of experiment can not explained by secondary antinuclei. 

\section{Antimatter from exotic sources }
Modern cosmology classifies as exotic sources of antimatter annihilation or decay of hypothetical particles of dark matter and evaporation of primordial black holes.
\subsection{Dark matter}
Dark matter makes up $\sim 85 \% $ of all matter in the universe. Its presence is implied in many astrophysical and astrological observations, including gravitational effects and large-scale structure formation. Such effects cannot be explained by the action of baryonic matter. Since dark matter has not yet been observed, if it exists, then it must interact through gravity with baryonic matter and radiation. The decay and annihilation of such particles can lead to the formation of antiparticles \cite{link16}.
\subsection{Primordial black hole}
Primordial black holes are a hypothetical type of black hole formed after the Big Bang. In the early Universe, high densities and inhomogeneous conditions could lead to gravitational collapse in dense regions, forming black holes. Ya. B. Zeldovich and I. D. Novikov in 1966 for the first time suggested the existence of such objects \cite{link18}. The theory of their origin was first deeply studied by S. Hawking in 1971 \cite{link19}.

Hawking showed that, due to quantum effects, black holes radiate like a black body with a temperature inversely proportional to the mass of the black hole. A physical understanding of the process can be obtained by imagining that particle-antiparticle radiation is emitted from beyond the event horizon. According to the modern concept, primordial black holes can also be sources of positrons and antiprotons \cite{link20}.
\newpage
 
\section{Globular clusters in the galactic halo}

A globular star cluster is a collection of stars that forms a spherical cluster rotating around the core of the Galaxy. Globular clusters are very closely connected by gravity, which gives them a spherical shape and a relatively high density of stars towards their centers. The name of this category of star clusters comes from the Latin globulus - a small sphere.

Globular clusters are located in the galactic halo and contain more stars and are much older than the less dense open clusters found in the galactic disk. Globular clusters are common, with about 150 globular clusters currently known in the Milky Way \cite{link39}.

Observations of globular clusters show that these stellar formations originate mainly in regions of effective star formation, where the interstellar medium is denser than normal star-forming regions. Currently, none of the known globular clusters show active star formation, they are free of gas and dust, and it is assumed that all the gas and dust were long ago either turned into stars or blown out of the cluster during the initial explosion of star formation. This is consistent with the opinion that globular clusters are the oldest objects in the Galaxy and were among the first clusters of stars \cite{link28}.

The trajectories of the globular clusters are eccentric and inclined to the plane of the galaxy. Orbiting the "outskirts" of a galaxy, globular clusters take several hundred million years to complete one orbit. Stars can reach a density of 100 to 1000 stars per cubic parsec in the center of a globular cluster. This is different from the density of stars around our Sun, which is estimated at about 0.14 stars per cubic parsec.

Globular clusters are usually made up of stars that have a low proportion of elements other than hydrogen and helium compared to stars like the Sun. The proportion of havier elements may indicate the age of a star, with older stars usually having lower metallicities \cite{link29}-\cite{link30}.
\newpage
\section{Discussion about the sourse function}

The paper considers a typical globular cluster, presumably consisting of anti-stars. As an example, we take one of the closest clusters - M4 in the Messier catalog (fugure \ref{m4parametrs}) (NGC 6121 in the new general catalog (NGC)). 
\begin{table}[H]
\caption{Parametrs of globular cluster M4 \cite{Heggie} }
\label{m4parametrs}
\begin{center}
\begin{tabular}{|c|c|c|}
\hline
Age, Gy & Distance from the Sun, kpc & Number of stars \\ 
\hline
12 & 1.72 & $8 \cdot 10^4$  \\
\hline
\end{tabular}
\end{center}
\end{table} Then we also assume that globular cluster M4 is a source of ${\overline{He}}^4$ in galactic cosmic rays.\\
\begin{center}
Three possible mechanisms for the injection of antihelium into cosmic rays from the globular cluster M4:
\end{center}
1. \textbf{\small Stationary outflow of matter from the surface of antistars } \\
If the diapason of propagation of antimatter from the globular cluster crosses the galactic disk, then the stellar wind will enter the disk, and then into the solar system. A stationary outflow of star matter in a cluster is considered for this. Stars are constantly losing part of their mass, so the concentration of particles from the entire globular cluster can be large. These are very low energies, a process of additional acceleration of particles is required to effectively overcome the solar magnetic field, but this effect is suppressed. In this case, we expect an energy $\sim$ MeV.\\
2. \textbf{\small Flares on antistars} \\
It is a known fact that active explosive processes occur on the Sun, accompanied by the acceleration of particles and the appearance of solar cosmic rays. We assume the existence of similar processes on antistars in a globular cluster. 

Particles from flares on antistars can receive higher energy ($\sim$ GeV), forming the antinuclear component of galactic cosmic rays. \bigskip \\
3. \textbf{\small Explosions of antisupernovae in a globular cluster of antistars} \\
Supernova explosions are the result of the evolution of stars, which is accompanied by the release of high energy up to $\sim 10^{51}$ erg.
The shell from the exploded anti-star propagates at high speed. Particles can acquire energy ($\sim10^{15}$ eV) as a result of various acceleration mechanisms on the supernova shell, accelerate and inject into cosmic rays. By analogy with the fact that stars are the source of particles in cosmic rays, antistars should be the main source of antiparticles in cosmic rays. Supernovae may be the most likely source of antinuclei in galactic cosmic rays. \\

\section{Results}

\subsection{ Supernova explosions}
The analysis begins with the most probable mechanism - antisupernova explosions, because the magnetic fields of the Galaxy prevent the penetration of low-energy antiparticles into the Galactic disk.

But it is also important to note that the frequency of the explosion of such supernovae is low against the background of outbursts of anti-stars and against the background of a constant outflow of stationary matter of anti-stars. The first two cases will be discussed later.\\

\subsubsection{Calculation of the energy density of antiparticles in cosmic rays}

\begin{figure}[H]
\centering
\includegraphics[scale=0.8]{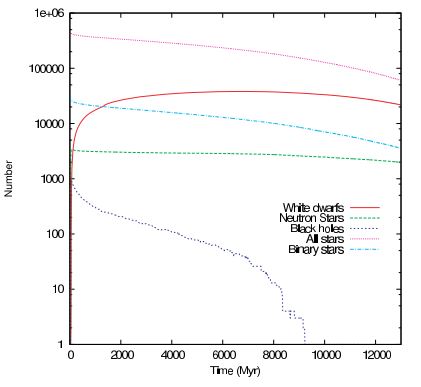}\\
\caption{ Change in the population of M4 in time}
\label{m4}
\end{figure}

Figure (\ref{m4}) shows a graph of the evolution of the population of the M4 cluster \cite{Montecarlo}. The graph shows the processes occurring in the early stages of the life of the cluster, the results of these processes can be compared with the present time. Let's pay attention to the number of neutron stars on the graph. Their number has not changed over 12 billion years. This means that about 12 billion years ago they could have formed as a result of the explosion of antisupernovae. This fact can be used to calculate the energy density of antiparticles in cosmic rays.
Using the formula for the energy density of cosmic rays of matter
\begin{equation}
\rho_{CR}= \frac{E_{sn}\dot N_{sn}t_{ret}}{V}
\end{equation}
$N_{sn}$ -- number of neutron stars in M4,
t-- cluster age,
$\dot N_{sn}$-- average supernova explosion frequency,
 $E_{sn} $ -- energy realized from supernova,
 $t_{ret}$-- cosmic ray lifetime, 
 V- volume of the region of propagation of cosmic rays (to calculate the volume, we considered a model of a cylinder with a height and radius of 30 kpc and 10 kpc, respectively. In order to consider not only the region of the disk, but also the halo of the Galaxy).

We present all the numerical values of these quantities in the form of a table figure (\ref{table}).
\begin{table}[H]
\caption{Table of numerical characteristics of quantities for calculating the energy density of antiparticles}
\label{table}
\begin{center}
\begin{tabular}{|c|c|c|c|c|c|}
\hline
$N_{sn}$ & t, Gy & $\dot{N_{sn}}$ & $E_{sn}$,erg & $t_{ret}$, myr & V,$kpc^3$ \\ 
\hline
12 & 1.72 & $8 \cdot 10^4$ & $10^{51}$ & $2 \cdot 10^{-5}$ & $3 \cdot 10^{3}$\\
\hline
\end{tabular}
\end{center}
\end{table}

the density using formula (1) and the values in the table:
 \begin{equation}
\rho_{\overline{CR}}\sim 10^{-4} \text{eV}/{\text{cm}}^3
\end{equation}
 For comparison, we present the value of the energy density of cosmic rays of matter:
 \begin{equation}
\rho\sim 1 \text{eV}/{\text{cm}}^3
\end{equation}
We also pay attention to the fact that the energy density for secondary antiprotons:
\begin{equation}
\rho_{\overline{p}}\sim 10^{-5} \text{eV}/{\text{cm}}^3
\end{equation}
The obtained value does not correspond to the established experimental data for the energy density of antiprotons. But given the fact that particles of cosmic rays pass through the magnetic fields of the Galaxy and lose some of the energy in them, it is necessary to consider in more detail the mechanism of motion of cosmic rays, which will be presented in the following part.

 \subsubsection{Particle motion in a Galaxy's magnetic field} 
In order for us to estimate the real fraction of particles from the initial flux that penetrates into the disk of the Galaxy, it is necessary to simulate the motion of particles in the magnetic field of the Galaxy.

\paragraph{Simulation of the magnetic field of the galaxy.}Based on the equations according to the data of \cite{MF}, we have compiled a function program, the input parameters of which are the coordinates in the Galaxy, and the output parameters are the components of the magnetic field in the Cartesian coordinate system.

The components of the magnetic field in a cylindrical coordinate system centered at the Galactic center taken from \cite{MF}:
\begin{displaymath}  
B_{\phi}= -\frac{B_{1}}{2R/R_{0}}\frac{z}{(z+z_0)}\bigl(\sqrt{(R/R_{0})^2+(z/z_{0})^2-z/z_{0}})\bigr)
\end{displaymath}
\begin{displaymath}
B_{R}=\frac{1}{2}B_{1}\frac{z_{0}^2}{(z+z)^2}tanh(R/R_o)
\end{displaymath}
\begin{displaymath}
B_{z}=\frac{0.1B_1z_0}{R_0}+\frac{1}{2}B_{1}\frac{z_{0}^2}{(z+z)^2}\biggl(\frac{tanh(R/R_o)}{R}+\frac{sech^2(R/R_0)}{R_0}\biggr)
\end{displaymath}

Where $ R_0 $ = 5 kpc and $ z_0 $ = 0.5pc are taken as scale lengths, and the $ B_1 $ parameter is free in \cite{MF}  and is determined by calibration, for example, by the magnetic field near the solar system according to the data of \cite{link38}.

\begin{figure}[htp]
\begin{minipage}[htp]{0.49\linewidth}
\center{\includegraphics[width=1.0 \linewidth]{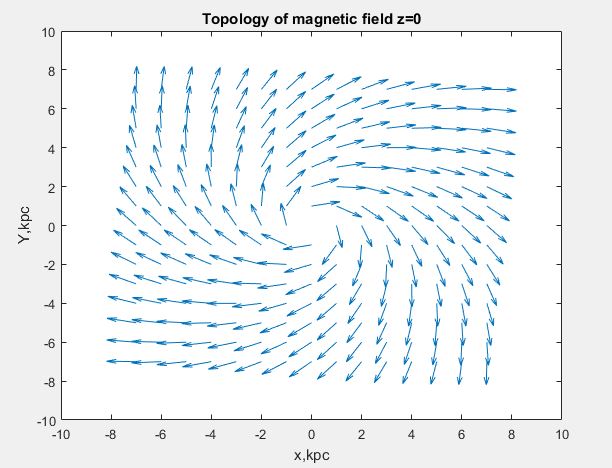} \\ a)}
\end{minipage}
\hfill
\begin{minipage}[htp]{0.49\linewidth}
\center{\includegraphics[width=1.0\linewidth]{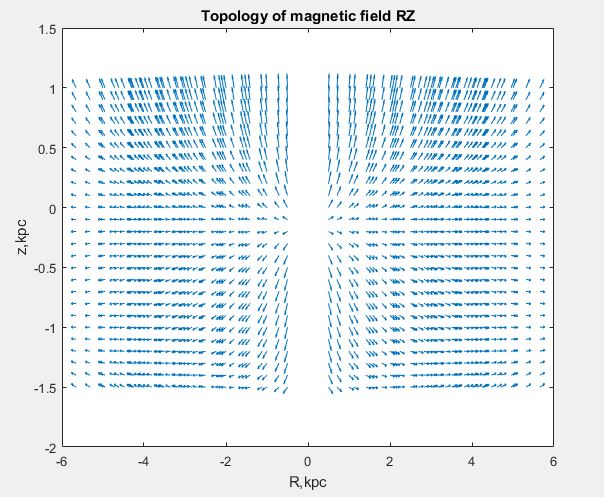} \\ b)}
\end{minipage}
\caption{\small Topology of the magnetic field of the Galaxy in the projection Z = 0 and in the projection RZ }
\label{topologies}
\end{figure}

We have constructed the topology of the Magnetic field of the Galaxy in the projection RZ and in the projection Z=0 (R and z are coordinates in a cylindrical coordinate system, a coordinate system centered at the Galactic center).

You can see in figure (\ref{topologies}a), that the magnetic field lines spiral out from the center of the Galaxy, this corresponds to the concept of the global magnetic field of the Galaxy in the plane of the galactic disk.
The figure (\ref{topologies}b) shows the vertical projection of the magnetic field of the Galaxy at y = 0, we see that the lines of force diverge in different directions according to the law determined by the equations from (\cite{MF}).
We have reproduced the magnetic field given in (\cite{MF}), and now we will simulate the propagation of particles in this magnetic field and will observe how particles are transported in our Galaxy.

\section{Conclusion}

	In this work, we considered the typical globular cluster M4, whose observed features can reproduce the expected properties of a globular cluster of anti-stars in the Galaxy. Based on the symmetry of the properties of matter and antimatter, we discuss the evolution of this GC and the mechanisms of injection of antimatter in CR.
	
We calculated the energy density of high energy antiparticles ejected by antimatter GC in cosmic rays, and also checked the operation of the program to simulate the propagation of these antiparticles in the magnetic field of the Galaxy.

      Further work is aimed at modeling the motion of particles in the magnetic field of the Galaxy, in order to estimate the minimum energy that a particle penetrating into a galactic disk should have. Implementation of our research program will help to obtain predictions of the expected flux of antinuclei as the signature of antimatter stars in our Galaxy.

\section*{Acknowledgements}
The research of MK was financially supported by Southern Federal University, 2020 Project № VnGr/2020-03-IF.
Research of AM was supported by the Ministry of Science and Higher Education of the Russian Federation under Project "Fundamental problems of cosmic rays and dark matter", No. 0723-2020-0040.


\begin{thebibliography}{99}




\bibitem{link1} P. A. M. Dirac: The quantum theory of the electron, Proc. Roy. Soc. (London) \textbf{A117}, 610—624 (1928).

\bibitem{newBook}
M. Y. Khlopov:   
{\em Fundamentals of Cosmoparticle Physics} CISP-Springer,Cambridge, UK, 2012.
\bibitem{DolgovAM}   A. D.  Dolgov: Matter and antimatter in the universe, Nucl. Phys. Proc. Suppl. {\bf 113} 40 (2002).

\bibitem{Oliva} Nicola Tomassetti, Alberto Oliva:  Secondary antinuclei from supernova remnants and background for dark matter searches,35th International Cosmic Ray Conference – ICRC2017, (2017).

\bibitem{link2} M.Yu. Khlopov: An antimatter globular cluster in our Galaxy - a probe for the origin of the matter, Gravitation and Cosmology , \textbf{4}, 69-72 (1998).

\bibitem{link3}A.D. Sakharov: Violation of CP-invariance, C-asymmetry and baryon asymmetry of the Universe, JETP Lett, \textbf{5}, 32 (1967).
\bibitem{link4} V.A. Kuzmin: CP violation and baryon asymmetry of the universe, JETP Lett, \textbf{12}, 228 (1970).
\bibitem{link5} V.M. Chechetkin, M.G. Sapozhnikov, M.Yu. Khlopov and Ya.B.Zeldovich: Astrophysical aspects of antiproton interaction with He (Antimatter in the Universe), Phys. Lett. \textbf{118B}, 359-362 (1982).
\bibitem{link6} V.M. Chechetkin, M.Yu. Khlopov and M.G. Sapozhnikov: Antiproton interactions with light elements as a test of GUT cosmologies., Rivista Nuovo Cimento, \textbf{5}, 1-80 (1982).
\bibitem{link7} A.D. Dolgov, A.F. Illarionov, N.S. Kardashev, I.D. Novikov, Cosmological model of a baryon island,  JETP, \textbf{67}, 1517-1524 (1988).
\bibitem{link8}M.Yu. Khlopov, S.G. Rubin, A.S. Sakharov:  Possible Origin of Antimatter Regions in the Baryon Dominated Universe., Phys.Rev.D \textbf{ 62}, 083505 (2000).
\bibitem{link9} M. Charlton, S. Eriksson, G. M. Shore: Fundamental Physics in Antihydrogen Experiments, 97-98 (2020).
\bibitem{link10} M.Yu. Khlopov, R.V. Konoplich, R. Mignani, et al.: Evolution and observational signature of diffused antiworld., Astroparticle Phys.,\textbf{12}, 367-372 (2000).

\bibitem{link13} I. Cholis, T. Linden: Anti-Deuterons and Anti-Helium Nuclei from Annihilating Dark Matter FERMILAB-PUB-20-021-A, (2020). 

\bibitem{link16} V. Trimble  Existence and Nature of Dark Matter in the Universe., Annual Review of Astronomy and Astrophysics.,\textbf{ 25}, 425-472 (1987).
\bibitem{link18} Ya.B. Zeldovich, I.D. Novikov: The Hypothesis of Cores Retarded During Expansion and the Hot Cosmological Model, Soviet Astronomy, \textbf{10}, 602 (1966).
\bibitem{link19} S. Hawking:  Gravitational collapsed objects of very low mass. Mon. Not. Roy. astron. Soc.,\textbf{152} , 75–78 (1971).

\bibitem{link20} W. Carroll Bradley,  A. Ostlie Dale:
{\em An Introduction to Modern Astrophysics}, Reading, MA: Addison-Wesley Publishing, 1996.
\bibitem{link39} http://gclusters.altervista.org/
\bibitem{link28} M. Paul: {\em Star Clusters. Encyclopedia of Astronomy and Astrophysics}, 2014.
\bibitem{link29}https://www.astro.keele.ac.uk/workx/globulars/globulars.html
\bibitem{link30} J. S. Kalirai, H. B.Richer: Star clusters as laboratories for stellar and dynamical evolution,	Royal society publishing, (2009).

\bibitem{Heggie} D. C. Heggie and M. Giersz: Modelling individual globular clusters, Cambridge University Press Access S246,\textbf{3}, 121-130 (2007).
\bibitem{Montecarlo} D. C. Heggie, M. Giersz:  Monte Carlo simulations of star clusters – V. The globular cluster M4,Monthly Notices of the Royal Astronomical Society 1, \textbf{388}, 429–443 (2008).

\bibitem{MF} C. J. Nixon, T. O. Hands: The origin of the structure of large–scale magnetic fields in disc galaxies Notices of the Royal Astronomical Society 3, \textbf{477}, 3539–3551 (2018).
\bibitem{link38} M. Opher, F. Alouani Bibi: A strong, highly-tilted interstellar magnetic field near the Solar System, Nature, \textbf{462}, 1036–1038 (2009).









\end{thebibliography}
\end{document}